\documentclass[conference,twocolumn]{IEEEtran}
\usepackage{times}
\usepackage[pdftex]{graphicx}
\usepackage{latexsym}
\usepackage{dsfont}
\usepackage{cite}
\usepackage{verbatim}
\usepackage{subfigure}
\usepackage[acronym,nogroupskip,nonumberlist,nopostdot]{glossaries}
\usepackage{xspace}
\usepackage{cite}
\usepackage{amsmath,amssymb,amsfonts}
\usepackage{algorithmic}
\usepackage{graphicx}
\usepackage{textcomp}
\usepackage{xcolor}
\usepackage{float}

\newcommand{\secref}[1]{{Sec.}~\ref{#1}}
\newcommand{\figref}[1]{{Fig.}~\ref{#1}}

\def\bb0{{\mathbb{0}}}

\def\bb{{\mathbf{b}}}

\def\b0{{\mathbf{0}}}

\def\sf0{{\mathsf{0}}}

\def\rm0{{\mathrm{0}}}

\def\b0{{\pmb{0}}} 

\usepackage{cite}
\usepackage{graphicx} 
\usepackage{amsmath}
\usepackage{amsfonts}
\usepackage{dsfont}
\usepackage{xspace}
\usepackage{xcolor}
\usepackage{svg}
\usepackage{tikz}

\newcommand\copyrighttext{%
  \footnotesize © 2026 IEEE. Personal use of this material is permitted. Permission from IEEE must be obtained for all other uses, in any current or future media, including reprinting/republishing this material for advertising or promotional purposes, creating new collective works, for resale or redistribution to servers or lists, or reuse of any copyrighted component of this work in other works.}

\newcommand\copyrightnotice{%
\begin{tikzpicture}[remember picture,overlay]
\node[anchor=south,yshift=10pt] at (current page.south) {\fbox{\parbox{\dimexpr\textwidth-\fboxsep-\fboxrule\relax}{\copyrighttext}}};
\end{tikzpicture}%
}

\definecolor{mygreen}{rgb}{0 0.5 0}

\newcommand{\topic}[1]{\textcolor{black}{#1}}
\newcommand{\dcomment}[1]{\textcolor{black}{#1}}

\newcommand{\trans}[1]{{#1}^\mathsf{T}}

\newcommand{\nServed}{\ensuremath{M}\xspace}
\newcommand{\nSlots}{\ensuremath{N}\xspace}
\newcommand{\slotIndex}{\ensuremath{n}\xspace}

\newcommand{\slotDuration}{\ensuremath{T_\text{s}}\xspace}
\newcommand{\totalNoise}{\ensuremath{N_0}\xspace}
\newcommand{\timeshare}{\ensuremath{\alpha}\xspace}

\newcommand{\txAntennaGain}{\ensuremath{\Gamma_\text{T}}\xspace}
\newcommand{\rxAntennaGain}{\ensuremath{\Gamma_\text{R}}\xspace}
\newcommand{\wavelength}{\ensuremath{\lambda}\xspace}
\newcommand{\userSchedVar}{\ensuremath{\mathds{1}(\userIndex,\slotIndex)}\xspace}
\newcommand{\contSchedVar}{\ensuremath{\tilde{\mathds{1}}(\userIndex,\slotIndex)}\xspace}
\newcommand{\scheduleSet}{\ensuremath{\mathcal{B}}\xspace}
\newcommand{\contScheduleSet}{\ensuremath{\tilde{\mathcal{B}}}\xspace}
\newcommand{\bsReceivePower}{\ensuremath{\text{P}^\text{B}_\slotIndex\xspace}}
\newcommand{\bsPos}{\ensuremath{\mathbf{x}^\text{B}}}
\newcommand{\gvSNR}{\ensuremath{\text{SNR}^\text{GV}_{\userIndex,\slotIndex}}\xspace}
\newcommand{\vbSNR}{\ensuremath{\text{SNR}^\text{VB}_\slotIndex}\xspace}
\newcommand{\gvSE}{\ensuremath{\text{SE}^\text{GV}_{\userIndex,\slotIndex}}\xspace}
\newcommand{\gvSEAvg}{\ensuremath{\text{SE}^\text{GV}_\slotIndex}\xspace}
\newcommand{\vbSE}{\ensuremath{\text{SE}^\text{VB}_\slotIndex}\xspace}
\newcommand{\slotLowerBound}{\ensuremath{\eta_\slotIndex}\xspace}
\newcommand{\slotLowerSet}{\ensuremath{\mathcal{S}}\xspace}
\newcommand{\gvLowerDist}{\ensuremath{d_{\userIndex,\slotIndex}}\xspace}
\newcommand{\gvLowerSet}{\ensuremath{\mathcal{G}}\xspace}
\newcommand{\vbLowerDist}{\ensuremath{d^\text{B}_\slotIndex}\xspace}
\newcommand{\vbLowerSet}{\ensuremath{\mathcal{V}}\xspace}
\newcommand{\bsHeight}{\ensuremath{b_\text{z}}\xspace}
\newcommand{\bandwidth}{\ensuremath{B}\xspace}

\newcommand{\userCenterVec}{\ensuremath{\overline{\mathbf{x}}^\text{G}}\xspace}
\newcommand{\distXMean}{\ensuremath{\mu_\text{x}}\xspace}
\newcommand{\distYMean}{\ensuremath{\mu_\text{y}}\xspace}
\newcommand{\distXDev}{\ensuremath{\sigma_\text{x}}\xspace}
\newcommand{\distYDev}{\ensuremath{\sigma_\text{y}}\xspace}
\newcommand{\userIndex}{\ensuremath{g}\xspace}
\newcommand{\uthUserPos}{\ensuremath{\mathbf{x}^\text{G}_\userIndex}}
\newcommand{\nUsers}{\ensuremath{G}\xspace}
\newcommand{\userTransmitPower}{\ensuremath{P^\text{G}_\text{tx}}\xspace}

\newcommand{\flightCenterVec}{\ensuremath{\mathbf{x}^\text{c}}\xspace}
\newcommand{\uavPos}{\ensuremath{\mathbf{x}^\text{V}_\slotIndex}\xspace}
\newcommand{\flightCenterX}{\ensuremath{c_\text{x}}\xspace}
\newcommand{\flightCenterY}{\ensuremath{c_\text{y}}\xspace}

\newcommand{\flightHeight}{\ensuremath{H}\xspace}
\newcommand{\flightRadius}{\ensuremath{r}\xspace}
\newcommand{\flightRadiusMin}{\ensuremath{\flightRadius_\text{min}}\xspace}

\newcommand{\flightSpeed}{\ensuremath{v}\xspace}
\newcommand{\flightSpeedMin}{\ensuremath{\flightSpeed_\text{min}}\xspace}
\newcommand{\flightSpeedMax}{\ensuremath{\flightSpeed_\text{max}}\xspace}
\newcommand{\uavTransmitPower}{\ensuremath{P^\text{V}
_\text{tx}}\xspace}
\newcommand{\uavReceivePower}{\ensuremath{\text{P}^\text{V}_{\userIndex,\slotIndex}}\xspace}

\date{June 2025}

\begin{document}

\title{Fixed-wing UAV relay optimization for coverage hole recovery} 

\author{\IEEEauthorblockN{Daniel T. Bonkowsky\IEEEauthorrefmark{1}, Ibrahim Kilinc\IEEEauthorrefmark{2}, and Robert W. Heath Jr.\IEEEauthorrefmark{2}}
	\IEEEauthorblockA{\IEEEauthorrefmark{1} Department of Computer Science and Engineering, University of California, San Diego, USA \\
		\IEEEauthorrefmark{2} Department of Electrical and Computer Engineering, University of California, San Diego, USA\\
		E-mail: \{dbonkowsky, ikilinc, rwheathjr\}@ucsd.edu
	}
}

\maketitle
\copyrightnotice
\thispagestyle{empty}
\pagestyle{empty}

\begin{abstract}



Unmanned aerial vehicles (UAVs) fill coverage holes as wireless relays during emergency situations. Fixed-wing UAVs offer longer flight duration and larger coverage in such situations than rotary-wing counterparts. Maximizing the effectiveness of fixed-wing UAV relay systems requires careful tuning of system and flight parameters. This process is challenging because factors including flight trajectory, timeshare, and user scheduling are not easily optimized. In this paper, we propose an optimization for UAV-based wireless relaying networks based on a setup which is applicable to arbitrary spatial user positions. In the setup, a fixed-wing UAV flies over a circular trajectory and relays data from ground users in a coverage hole to a distant base station (BS). Our optimization iteratively maximizes the average achievable spectral efficiency (SE) for the UAV trajectory, user scheduling, and relay timeshare. The simulation results show that our optimization is effective for varying user distributions and that it performs especially well on distributions with a high standard deviation.
\end{abstract}



\section{Introduction} \label{sec:introduction}

UAVs can be an essential tool for providing coverage during a disaster when the conventional cellular infrastructure becomes inoperable. A significant amount of work has considered the role of rotary-wing UAVs acting as relays between a functioning BS tower and ground-based users \cite{OpportunitiesAndChallenges,TrajOptimization5G}. Unfortunately, there is much less work on fixed-wing UAVs to solve the same challenge. \dcomment{Fixed-wing UAVs present a more challenging optimization problem as they have a greater turning radius than rotary-wing UAVs and must fly above a minimum velocity to remain airborne.} Fixed-wing UAVs have benefits of higher flight altitude with larger coverage, longer operation duration, endurance in harsh weather conditions, energy efficiency, and payload capacity \cite{FlightPerformance,MaximumEndurance,OpportunitiesAndChallenges,UAVsInDisasters}. Therefore, fixed-wing UAVs represent a promising tool as an aerial relay for emergency situations to enable connectivity. In this paper, we consider a fixed-wing UAV relay system for distant users in a coverage hole and propose an optimization to maximize the achievable spectral efficiency of the system.

\topic{Relay system optimizations are primarily characterized by the choice of optimization parameters and the setup of the system model.} Prior work has demonstrated that system throughput can be maximized by optimizing trajectory \cite{EnergyEfficiencyMaximization,3-DTrajOptimization,MobileUsers}, transmission power \cite{EnergyEfficiencyMaximization,MobileUsers}, user scheduling \cite{CyclicalMultipleAccess,MobileUsers}, and relaying timeshare \cite{WirelessRelayNetwork}. A fixed-wing UAV relay in \cite{EnergyEfficiencyMaximization} provides a line-of-sight link for edge users in a cell with a BS. Users are, however, assumed to be directly below the circular UAV trajectory. Multiple fixed-wing UAVs act as relays to aid an overloaded BS, serving users located in a ring around the BS in \cite{3-DTrajOptimization}. A fixed-wing UAV in \cite{MobileUsers} acts as a BS for a mobile group of users, flying with them as they move. A fixed-wing UAV in \cite{CyclicalMultipleAccess} serves as an aerial BS for users that are assumed to be evenly spaced along a straight line. A fixed-wing UAV relays between two stationary ground terminals in \cite{WirelessRelayNetwork}, flying on a constant circular path. In real emergency scenarios, users are distributed according to population density, terrain accessibility, and the nature of the disaster, none of which conform to these idealized geometries. The system model in \cite{MobileUsers} addresses a general distribution of users, but the UAV acts as an aerial BS with a single-hop link from the ground users. Furthermore, the prior work lacks consideration of relay systems \cite{MobileUsers,CyclicalMultipleAccess}, and realistic user distributions \cite{EnergyEfficiencyMaximization,3-DTrajOptimization,WirelessRelayNetwork} in coverage holes. Therefore, our work of optimizing fixed-wing UAV relay systems to support distant coverage holes during emergencies is especially relevant.

\topic{We propose a fixed-wing UAV relay system serving distant users in coverage holes without assuming a specific user distribution.} We develop an iterative optimization to tune circular trajectory, user scheduling, and time allocation to maximize connectivity. The UAV follows a circular flight path and uses decode-and-forward relaying between ground users and a stationary BS. The relay system has a slotted communication model in which groups of users are scheduled at each time slot. There is no restriction on the location of the ground users within the coverage dead zone, thus our setup is applicable to arbitrary real-world scenarios. Based on this system, we formulate an optimization of the flight trajectory, user scheduling, and relay timeshare, and solve the resulting non-convex problem via iterative optimization. Through numerical simulation, we examine the effects of UAV transmission power, flight altitude, and user distribution on the optimization. Results show that the proposed optimization algorithm is resilient to increasing standard deviation of the user distribution, adapts to distribution characteristics, and provides gains over a baseline model. 


\section{System model} \label{sec:sys-model}
In this section, we describe the network topology of the fixed-wing UAV relay system and slotted communication model. Then, we define the equations used to model signal transmission.
\begin{figure}
    \centering
    \includegraphics[width=0.9\linewidth]{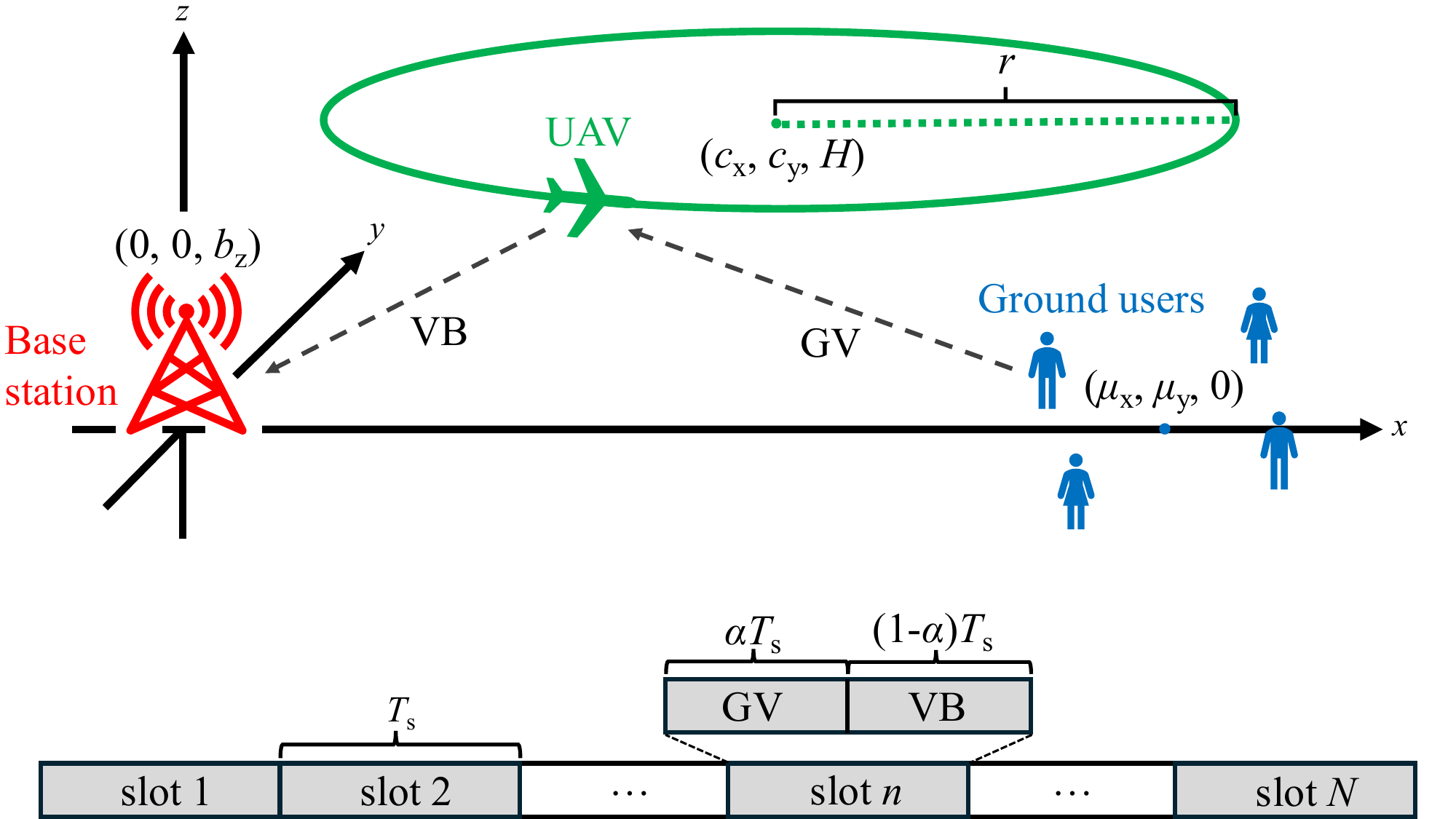}
    \caption{The system and communication model of the UAV-enabled wireless relay system. The UAV flies on a circular trajectory at a fixed altitude and serves users in a distant coverage dead zone by connecting them with a BS located near the origin. We use a slotted communication model, with multiple users served per time slot. Each time slot is divided between the user-to-UAV link (denoted GV) and UAV-to-BS link (denoted VB).}
    \label{fig:system-model}
\end{figure}

\subsection{Network topology}
\topic{We consider a wireless system with a fixed-wing UAV relay that serves ground users in a dead zone by connecting them with a distant BS.} There is assumed to be no direct connectivity between the users and the BS due to mountains, buildings, or other obstacles that may be present in disasters and emergencies. The UAV relays with a decode-and-forward architecture, and is equipped with a data buffer large enough to handle data from multiple users. The decoding and re-encoding process is assumed to be instantaneous, and thus the UAV is always either receiving from users or transmitting to the BS. We only consider one-way relaying from user to UAV to BS, but similar analysis can be applied to relaying in the opposite direction. We use 3D Cartesian coordinates as a frame of reference, where the origin lies at $(0, 0, 0)$, the xy-plane represents the ground, and the z-axis represents altitude. \figref{fig:system-model} shows an overview of the proposed system.

\topic{The physical model of the system is characterized by its three main components: the BS, the ground users, and the UAV.} The BS is modeled as a point at the location $\bsPos = \trans{[0, 0, \bsHeight]}$. There are \nUsers users within the dead zone, which is centered at the location $\userCenterVec = \trans{[\distXMean, \distYMean, 0]}$. The $\userIndex$th user has coordinates denoted $\uthUserPos=\trans{[x_\userIndex, y_\userIndex, 0]}$, which can fall anywhere within the coverage dead zone. The UAV relays between the users and BS from a circular path located at a fixed altitude. We choose a circular path because it is simple and effective in terms of providing connectivity \cite{NovelMobilityModel,EnergyEfficiencyMaximization,TrajOptimization5G}. The flight path has radius \flightRadius and center point $\flightCenterVec = \trans{[\flightCenterX, \flightCenterY, \flightHeight]}$, where \flightHeight is constant. We assume the UAV flies with constant speed \flightSpeed, which is subject to $\flightSpeedMin < \flightSpeed < \flightSpeedMax$. Here, \flightSpeedMin is the minimum speed of the fixed-wing UAV required to remain airborne, and \flightSpeedMax is determined by the build of the UAV and the radius of the flight path \cite{AerodynamicEfficiency}.

\topic{The communication system is a slotted model, with distinct users served in each time slot.} There are \nSlots time slots, indexed by $\slotIndex=1,\dots,\nSlots$. Each has length \slotDuration, which is short enough that the UAV can be considered stationary over the time slot \cite{JointTrajectoryCommunicationDesign,3DTrajAndTransmitPower,BackhaulAwareTrajOpt,JointUserSched,Opt3DTrajDesign}. The UAV's trajectory can thus be considered a discrete sequence of positions, each one a function of the flight path radius \flightRadius and the center point of the flight path \flightCenterVec. Then the UAV's position during time slot \slotIndex is given as
\begin{equation} \label{eq:uav-position}
\begin{split}
    \uavPos&(\flightRadius,\flightCenterVec) =\\&\trans{\left[\flightCenterX + \flightRadius \cos\left(\frac{2\pi \slotIndex}{\nSlots}\right),\flightCenterY + \flightRadius \sin\left(\frac{2\pi \slotIndex}{\nSlots}\right),\flightHeight\right]}.
\end{split}
\end{equation}
While the starting point of the UAV's trajectory determines all its subsequent positions, it has negligible impact on the system performance, since we always consider an integer number of rotations. The UAV serves $\nServed << \nUsers$ users per time slot using frequency-division multiple access \dcomment{\cite{MobileUsers, Opt3DTrajDesign}}, and bandwidth is uniformly allocated to ensure fair frequency assignment. To model user scheduling, we use the binary decision variable \userSchedVar, where $\scheduleSet = \{\userSchedVar \mid \forall \slotIndex\in1,\dots,\nSlots,\forall\userIndex\in1,\dots,\nUsers\}$ represents all user assignments in every time slot. If user \userIndex is served in time slot \slotIndex, then $\userSchedVar = 1$, and otherwise $\userSchedVar = 0$. In each time slot, users transmit their data to UAV for the duration $T_0$ before the UAV forwards this data to the BS for the duration $\slotDuration - T_0$. We define $\timeshare = T_0/\slotDuration$ as the proportion of each time slot reserved for user-to-UAV transmission, and thus $1-\timeshare$ is the proportion of each time slot reserved for UAV-to-BS transmission.

\subsection{Received signal model}
\topic{Based on the relay model, we derive the equations for computing the achievable rate.} We assume that both the user-to-UAV link (denoted GV) and the UAV-to-BS link (denoted VB) are line-of-sight, and the channels are expressed using a free-space propagation model \cite{heath2018foundations}. \dcomment{We assume that Doppler is already corrected as part of carrier frequency offset estimation.} Let the users have fixed transmission power \userTransmitPower, and let the UAV have transmission power \uavTransmitPower. Let \wavelength denote the signal wavelength, let \txAntennaGain denote the antenna gain at the transmitter, let \rxAntennaGain denote the antenna gain at the receiver, and let \bandwidth denote the frequency bandwidth. Let $\|\cdot\|$ denote the $2$-norm throughout the manuscript. Using free-space propagation, the power of the signal received by the UAV from the $\userIndex$th user in the $\slotIndex$th time slot is $\uavReceivePower(\flightRadius,\flightCenterVec) = \userTransmitPower\txAntennaGain\rxAntennaGain\left(\frac{\wavelength}{4\pi||\uavPos-\uthUserPos||}\right)^2$, and the power of the signal received by the BS in the $\slotIndex$th time slot can be expressed as $\bsReceivePower(\flightRadius,\flightCenterVec)=\uavTransmitPower\txAntennaGain\rxAntennaGain\left(\frac{\wavelength}{4\pi||\uavPos||}\right)^2$. Let \totalNoise be the noise power spectral density. In the user-to-UAV link, there are \nServed users transmitting simultaneously to the UAV with uniform band allocation so the channel for a single user is subject to noise power $\totalNoise\bandwidth/\nServed$. Thus, the signal-to-noise ratio (SNR) of the GV link for the $\userIndex$th user during the $\slotIndex$th time slot can be expressed as $\gvSNR = \frac{\uavReceivePower}{\totalNoise\bandwidth/\nServed}$ and the SNR of the VB link during the $\slotIndex$th time slot can be expressed as $\vbSNR = \frac{\bsReceivePower}{\totalNoise\bandwidth}$, where $\flightRadius,\flightCenterVec$ are omitted for notational convenience. Assuming Gaussian signaling under Gaussian noise, the achievable spectral efficiency of the GV link for the $\userIndex$th user during the $\slotIndex$th time slot can be expressed $\gvSE(\flightRadius,\flightCenterVec) = \log_2(1 + \gvSNR)$ \cite{heath2018foundations}. The achievable spectral efficiency of the VB link during the $\slotIndex$th time slot can be expressed $\vbSE(\flightRadius,\flightCenterVec)=\log_2(1+\vbSNR)$.

\topic{The performance of the wireless relay system is measured by the average achievable spectral efficiency per time slot per user}. We define $\gvSEAvg(\flightRadius,\flightCenterVec,\scheduleSet)=\frac{1}{\nServed}\sum_{\userIndex=1}^\nUsers \userSchedVar\gvSE$ as average per-user spectral efficiency in the GV link. We assume perfect channel knowledge, and that all users are perfectly synchronized over time slots. Let $\min(a,b)$ be the function of the minimum of $a$ and $b$. The average achievable SE of the relay system is given as
\begin{equation} \label{eq:mean-se}
\begin{split}
    \overline{\text{SE}}&(\timeshare,\flightRadius,\flightCenterVec, \scheduleSet) =\\&\frac{1}{\nSlots}\sum_{\slotIndex=1}^\nSlots \min\left(\alpha\cdot\gvSEAvg\;\mathbf{,}\;(1-\timeshare)\cdot\vbSE\right).
\end{split}
\end{equation}
The argument of $\overline{\text{SE}}(\cdot)$ metric in \eqref{eq:mean-se} indicates the optimization variables. All other variables introduced so far are fixed.
\section{Problem formulation} \label{sec:prob-formulation}
\topic{The objective is to maximize achievable SE per-user per-time slot of the relay system expressed in \eqref{eq:mean-se}.} The optimization involves the variables of timeshare proportion \timeshare, the user scheduling set $\scheduleSet$ defined for $\nSlots$ time slots, and circular trajectory center $\flightCenterVec$ and radius $\flightRadius$. The optimization problem is a non-convex and non-linear program \cite{ConvexOptimization}, and optimizing all parameters jointly is not tractable. Instead, we propose an iterative maximization approach by decomposing the problem into timeshare optimization, scheduling optimization, and trajectory optimization, where each subproblem is better behaved for optimization. The minimum function in equation \eqref{eq:mean-se} is problematic for optimization, since it is non-convex and non-differentiable in its arguments \cite{ConvexOptimization}. To address this, we define a lower-bound \slotLowerBound on the achievable SE in the $\slotIndex$th time slot, subject to constraints $\slotLowerBound \leq \timeshare \gvSEAvg$ and $\slotLowerBound \leq (1-\timeshare)\vbSE$, with $\slotLowerSet = \{\slotLowerBound \mid \forall \slotIndex \in 1, \dots, \nSlots\}$ defined as the lower-bound on the achievable spectral efficiency in each time slot. These constraints allow us to maximize the average spectral efficiency while reformulating each subproblem into an efficiently solvable form.

\subsection{Timeshare optimization}
\topic{The first subproblem in the iterative maximization is to optimize the timeshare variable \timeshare for given trajectory parameters \flightRadius, \flightCenterVec, and scheduling set  \scheduleSet.} The problem can be formulated into a linear program given as

\begin{subequations} \label{eq:timeshare-optimization}
\begin{align}
    \max_{\timeshare, \slotLowerSet} \quad & \frac{1}{\nSlots}\sum_{\slotIndex=1}^\nSlots \slotLowerBound\\
    \text{subject to } & \slotLowerBound \leq \timeshare \gvSEAvg, \forall \slotIndex\\
    & \slotLowerBound \leq (1 - \timeshare) \vbSE, \forall \slotIndex\\
    & 0 \leq\timeshare\leq1.
\end{align}
\end{subequations}
The optimization problem has a linear objective function and affine constraints. Thus, it can be optimized with solvers such as MOSEK, which guarantees solutions that are numerically primal- and dual-feasible \cite{mosek2025pythonapi}.

\subsection{Scheduling optimization}
\topic{For a given timeshare proportion \timeshare and trajectory variables \flightRadius and \flightCenterVec, the user scheduling set \scheduleSet is optimized using a similar method to timeshare optimization}. The binary user-scheduling variables turn the problem into a mixed-integer optimization, which is difficult to solve efficiently \cite{MixedIntOpt}. To address this, we relax the binary scheduling variables to be continuous, constrained by $0\leq\contSchedVar\leq1$, where $\contScheduleSet = \{\contSchedVar \mid \forall n, u\}$ is the set of all relaxed scheduling variables. We formulate additional constraints to allow at most \nServed users to be scheduled in each time slot and each user to be scheduled at most $\frac{\nSlots\nServed}{\nUsers}$ times over all time slots. Then the optimization problem can be written as
\begin{subequations} \label{eq:schedule-optimization}
\begin{align}
    \max_{\contScheduleSet, \slotLowerSet} \quad & \frac{1}{\nSlots}\sum_{\slotIndex=1}^\nSlots \slotLowerBound\\
    \text{subject to } & \slotLowerBound \leq \alpha\gvSEAvg,\forall\slotIndex \\
    & \slotLowerBound \leq (1 - \timeshare) \vbSE, \forall\slotIndex \\
    & 0 \leq \contSchedVar \leq 1, \forall \userIndex, \slotIndex \\
    & \sum_{\userIndex=1}^\nUsers \contSchedVar \leq \nServed,\forall \slotIndex \\
    & \sum_{\slotIndex=1}^\nSlots \contSchedVar \leq \frac{\nSlots\nServed}{\nUsers},\forall \userIndex.
\end{align}
\end{subequations}
The optimization problem is a linear program with a linear objective function and affine constraints, which can be efficiently optimized with solvers such as MOSEK. Upon completion of the scheduling optimization at each iteration, the binary scheduling set \scheduleSet is reconstructed from \contScheduleSet by assigning 1 to the \nServed greatest values in each time slot and 0 to the remaining. This process allows an efficient and realistic user scheduling.

\subsection{Trajectory optimization}
\topic{The trajectory optimization problem is not directly solvable, since the spectral efficiency functions \gvSE and \vbSE are non-convex with respect to \flightRadius and \flightCenterVec.} To address this, we employ a successive convex approximation (SCA) approach \cite{3-DTrajOptimization,JointTrajectoryCommunicationDesign,JointUserSched}. SCA handles non-convex optimization problems through an iterative procedure that successively solves tractable subproblems, each obtained by substituting the non-convex functions with simpler convex surrogates. Let us consider optimizing the non-convex function $V(\mathbf{x})$ over the convex set $\mathcal{X}$ using surrogate function $\tilde{V}(\mathbf{x}\mid\mathbf{x}^i)$ through SCA. The $i$th iteration of SCA consists of solving a subproblem of the form $\mathbf{x}^{i+1} \in \arg\max\tilde{V}(\mathbf{x}\mid\mathbf{x}^i)$, where $\mathbf{x}^i$ is the solution found during iteration $i-1$. To guarantee a stable SCA convergence, each surrogate function must be continuous on $\mathbf{x}\in \mathcal{X}$, they must be a global lower-bound of the original function, and the gradient and function value at $\mathbf{x}^i$ must match the original function \cite{SCABook}.

\topic{To reformulate trajectory optimization into an SCA problem, it is necessary to construct surrogate functions for \gvSE and \vbSE.} To do this, we first define constants $A_\text{G} = \frac{\userTransmitPower \txAntennaGain \rxAntennaGain \nServed}{\totalNoise\bandwidth}\left(\frac{\wavelength}{4\pi}\right)^2$ and $A_\text{B} = \frac{\uavTransmitPower \txAntennaGain \rxAntennaGain}{\totalNoise\bandwidth} \left(\frac{\wavelength}{4\pi}\right)^2$, auxiliary variables $\gvLowerDist = ||\uavPos - \uthUserPos||^2$ and $\vbLowerDist = ||\uavPos||^2$, and sets $\gvLowerSet = \{\gvLowerDist\mid\forall\userIndex,\slotIndex\}$ and $\vbLowerSet = \{\vbLowerDist \mid \forall \slotIndex\}$. Next, we redefine spectral efficiency functions \gvSE and \vbSE using these auxiliary variables as 
\begin{equation}
    \gvSE(\gvLowerDist) = \log_2\left(1 + \frac{A_\text{G}}{\gvLowerDist}\right)
\end{equation}
 and 
 \begin{equation}
     \vbSE(\vbLowerDist) = \log_2\left(1 + \frac{A_\text{B}}{\vbLowerDist}\right),
 \end{equation} which are equivalent to the original \gvSE and \vbSE for any \flightRadius and \flightCenterVec, and convex with respect to \gvLowerDist and \vbLowerDist. For SCA maximization, it is necessary that surrogate functions are concave with respect to their parameters, thus, the surrogate functions we propose are the first-order Taylor expansions of \gvSE and \vbSE \cite{ConvexOptimization}. This is a reasonable choice, because any convex function is globally lower-bounded by its first-order Taylor expansion \cite{ConvexOptimization}, and because linear functions can be efficiently optimized. Lastly, any equality constraints must be affine for an optimization problem to be convex, so the equality constraints on \gvLowerDist and \vbLowerDist need to be addressed \cite{ConvexOptimization}. We relax the equality constraints on \gvLowerDist and \vbLowerDist, replacing them with inequality constraints as the upper bounds for the squared user-to-UAV and UAV-to-BS distances. Optimality is preserved under this relaxation, as the maximization of \gvSE and \vbSE inherently minimizes \gvLowerDist and \vbLowerDist. This ensures the inequality constraints are satisfied as equality at the optimal solution. Thus, the fully formulated subproblem being solved in the $i$th iteration of SCA can be written as
\begin{subequations} \label{eq:trajectory-optimization}
\begin{align}
    \max_{\flightRadius,\flightCenterVec,\gvLowerSet,\vbLowerSet,\slotLowerSet} \quad &\frac{1}{\nSlots}\sum_{\slotIndex=1}^\nSlots \slotLowerBound\\
    \begin{split} \label{eq:ua-constraint}
        \text{subject to } &\slotLowerBound \leq \frac{\timeshare}{\nServed}\sum_{\userIndex=1}^\nUsers \userSchedVar\cdot \\
        &\;\;\left[\gvSE\left({\gvLowerDist}^{(i)}\right) + \right.\\ &\;\;\left. {\gvSE}'\left({\gvLowerDist}^{(i)}\right)\left(\gvLowerDist - {\gvLowerDist}^{(i)}\right)\right], \forall\slotIndex
    \end{split}\\
    \begin{split} \label{eq:ab-constraint}
        &\slotLowerBound  \leq (1-\timeshare)\cdot\\
        &\;\;\left[\vbSE\left({\vbLowerDist}^{(i)}\right) + \right.\\ &\;\;\left.{\vbSE}'\left({\vbLowerDist}^{(i)}\right)\left(\vbLowerDist - {\vbLowerDist}^{(i)}\right)\right], \forall\slotIndex
    \end{split}\\
    &\gvLowerDist \geq ||\uavPos-\uthUserPos||^2, \forall\userIndex,\slotIndex\label{eq:snk-constraint}\\
    &\vbLowerDist \geq ||\uavPos||^2,\forall\slotIndex, \label{eq:sbn-constraint}
\end{align}
\end{subequations}
where ${\gvLowerDist}^{(i)}$ and ${\vbLowerDist}^{(i)}$ are the values of \gvLowerDist and \vbLowerDist found from the $(i-1)$th SCA iteration. The objective function is linear, constraints \eqref{eq:ua-constraint} and \eqref{eq:ab-constraint} are affine, and constraints \eqref{eq:snk-constraint} and \eqref{eq:sbn-constraint} are convex. Therefore, the problem solved during each iteration is a convex optimization problem, which can be efficiently solved with guarantees as described in the timeshare optimization subsection. Our formulation leads to a stable convergence because \gvLowerDist and \vbLowerDist are bounded from below, so cannot be arbitrarily small. Values for \gvLowerDist and \vbLowerDist also cannot be arbitrarily large, since \gvSE and \vbSE are decreasing over \gvLowerDist and \vbLowerDist.

\begin{figure}[H]
    \centering
    \includegraphics[width=0.9\linewidth]{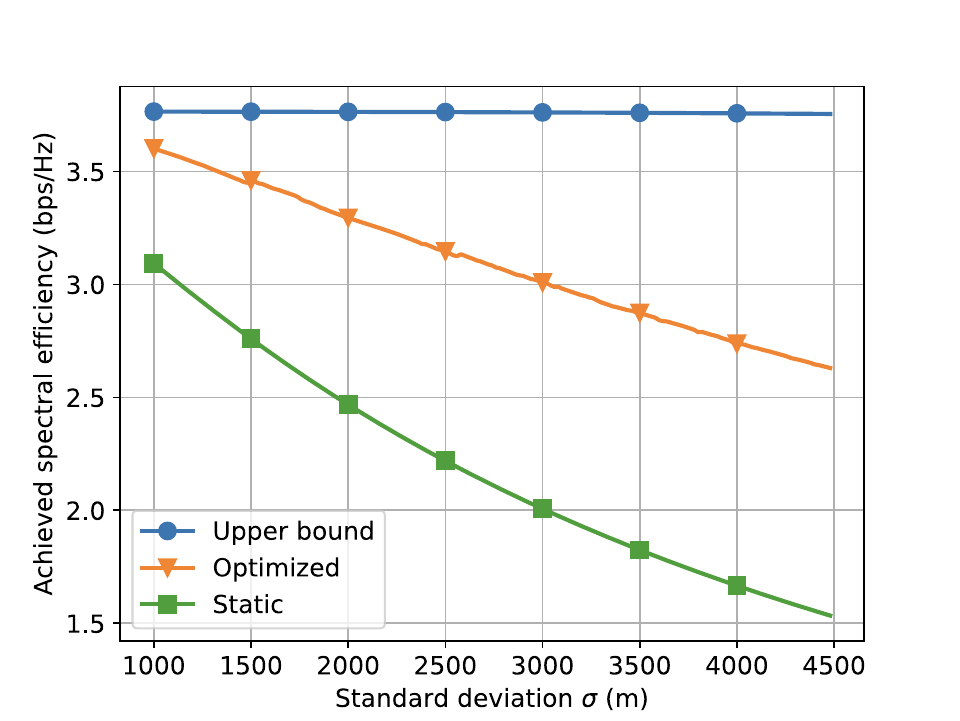}
    \caption{Impact of the standard deviation of the user distribution on the achieved SE of the system. The upper bound performs well regardless of standard deviation, because it is optimized per-user. Comparing the optimized model with the static model, which represents a reasonable heuristic, shows that our optimization is more robust to standard deviation increases.}
    \label{fig:variance-vs-gain}
\end{figure}

\section{Numerical results} \label{sec:num-results}
\topic{In this section, we present numerical results by focusing on a specific network topology.} The simulation involves $\nUsers=20$ ground users, with $\nServed=2$ users served concurrently per time slot. A complete flight path traversal is divided into $\nSlots=500$ time slots. The UAV maintains a circular trajectory at an altitude of $\flightHeight=1000$m with a minimum radius of $\flightRadiusMin=500$m. The user transmission power is set to $\userTransmitPower=0.01$W \cite{DronePublicSafety}, and the UAV transmission power is set to $\uavTransmitPower=10$W \cite{LTEUAS} unless stated otherwise. The spatial user distribution follows a bivariate normal distribution with means $\distXMean, \distYMean$ and standard deviations $\distXDev, \distYDev$, following prior work that has shown this to be a realistic model for user hotspots \cite{NormalUserModel}.

\topic{To evaluate our optimization, we compare its performance against two baselines: an ideal upper bound and a static, heuristic-based model.} The upper bound represents the average SE achieved by a rotary-wing UAV serving users in a round-robin fashion. For each user, it hovers at the optimal location with optimal timeshare, optimized by the methods described in \secref{sec:prob-formulation}. It represents an unattainable upper-bound in practice since the rotary-wing UAV cannot hover instantaneously at each time slot. The static baseline uses random user scheduling and a fixed circular trajectory centered at $\flightCenterVec = \trans{[\distXMean, \distYMean, \flightHeight]}$ with $\flightRadius=\flightRadiusMin$. To calculate the relay timeshare \timeshare for the static baseline, we model the distribution as a single user located at the center of the dead zone, and the UAV as a stationary relay located at \flightCenterVec. Let $\text{SE}_1$ be the spectral efficiency of the single user-to-relay link, and let $\text{SE}_2$ be the spectral efficiency of the relay-to-BS link. The optimal relay timeshare for the static baseline can be calculated as $\timeshare = \frac{\text{SE}_2}{\text{SE}_1+\text{SE}_2}$ \cite{PetersEtAlRelayArchitectures3GPPLTEAdvanced2009}. \dcomment{Prior work either does not address scenarios with varying user distributions \cite{EnergyEfficiencyMaximization,3-DTrajOptimization,WirelessRelayNetwork} or does not consider relay systems \cite{MobileUsers,CyclicalMultipleAccess}, which makes direct comparison challenging. Nevertheless, we adapt the rotary-wing approach from prior work to establish a meaningful upper-bound baseline and employ the optimal timeshare strategy from \cite{PetersEtAlRelayArchitectures3GPPLTEAdvanced2009} as a static baseline for our setting.}

\subsection{Impact of variance on spectral efficiency}
\topic{The plot in \figref{fig:variance-vs-gain} shows the impact of user variance on the spectral efficiency achieved by the system for two baselines and our method.} \dcomment{The plots were generated using 3000 Monte Carlo runs.} The results show that with increasing standard deviation, the upper bound is unchanged, while the optimized and static trajectory suffer spectral efficiency losses. The upper bound is optimized per-user, so it is expected that increasing the spread of users has little effect on overall SE. Comparing our approach with the static baseline shows the efficacy of our optimization. As users become more spread out, the UAV can respond and alter its trajectory to serve users better. Accordingly, the optimized system has a smaller decrease in spectral efficiency performance than the static system as the standard deviation increases. Consequently, our optimization becomes more relevant as the users are more spread out.

\subsection{Impact of UAV transmission power on spectral efficiency}
\begin{figure}
    \centering
    \includegraphics[width=0.9\linewidth]{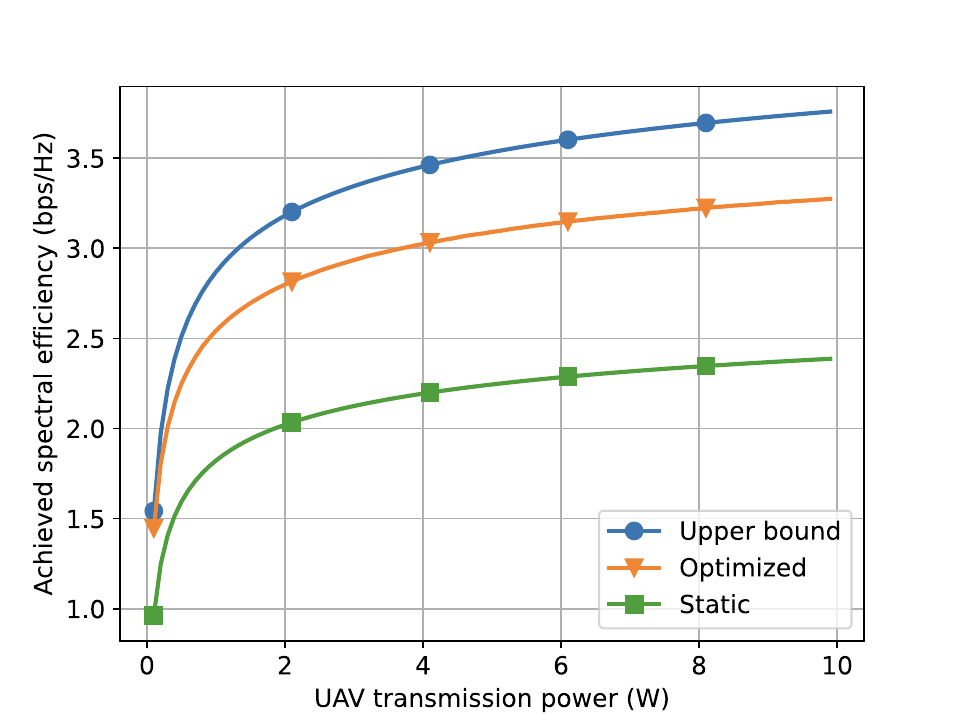}
    \caption{The effect of UAV transmission power on the SE achieved in the system. As UAV transmission power increases, the throughput bottleneck becomes the transmission power of the users, and the achievable spectral efficiency levels off. The static model cannot adapt to utilize the increase in UAV transmission power and levels very quickly. The optimized model can take advantage of increased transmission power by flying closer to users, mitigating the user-to-UAV limitation.}
    \label{fig:dist-vs-gain}
\end{figure}
\topic{\figref{fig:dist-vs-gain} illustrates the effect that UAV transmission power has on the SE.} Increasing UAV transmission power yields significant SE gains initially; however, the performance subsequently saturates, which reveals the inherent bottleneck of wireless relay systems. Total throughput is given by the minimum of the user-to-UAV and UAV-to-BS links, so even a system with a powerful UAV transmitter will perform poorly if user transmission power is weak. The optimized trajectory performs much better than static baseline, especially for greater UAV transmission power. This demonstrates the flexibility of our optimization, that is able to use high transmission power to achieve SE gains over the static baseline.

\subsection{Impact of UAV transmission power on optimal radius}
\begin{figure}
    \centering
    \includegraphics[width=0.9\linewidth]{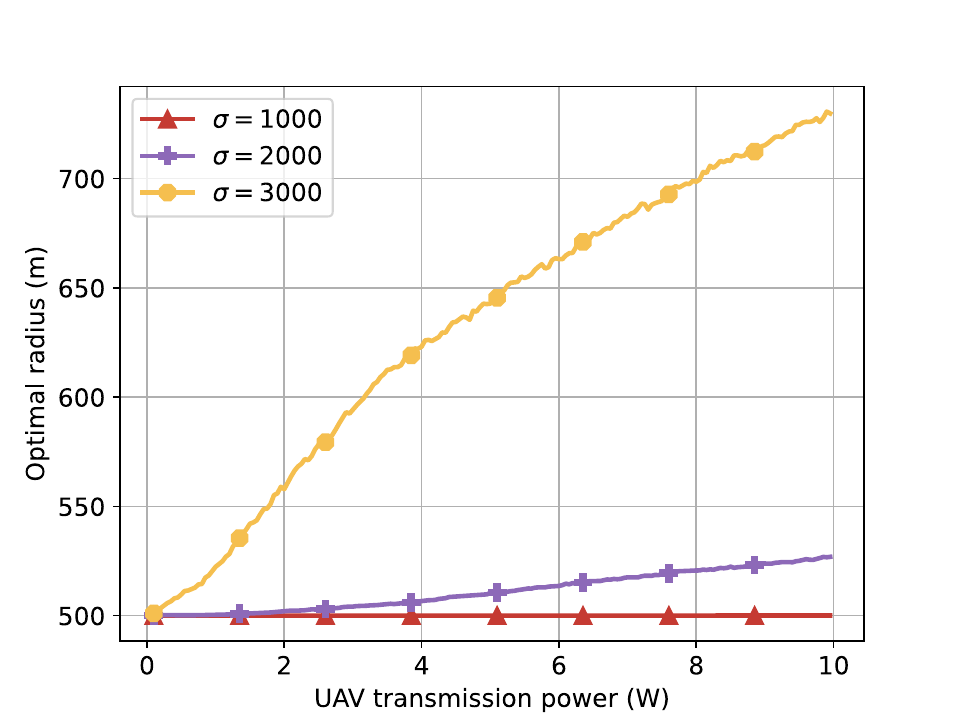}
    \caption{Impact of the UAV transmission power on the optimized radius. Optimal radius is positively correlated with transmission power, because greater transmission power gives the UAV more freedom to travel far from the BS and closer to ground users. This freedom is more pronounced in distributions with a larger standard deviation, because a larger radius is required to serve users when their spread is greater.}
    \label{fig:patx-vs-r}
\end{figure}
\topic{Next, we consider how the UAV’s transmission power affects the optimal radius of the flight path.} We present results for user distributions with standard deviations $\distXDev = \distYDev =$ 1000m, 2000m, and 3000m. \figref{fig:patx-vs-r} demonstrates that as UAV transmission power increases, the optimal flight path radius increases with it. For a UAV with greater transmission power, it is optimal to fly further from the BS and closer to ground users, who have a weaker transmission power. For distributions with a larger standard deviation, flying closer to users requires a larger radius, while distributions with a small standard deviation can be effectively served by a UAV circling with a small radius. \dcomment{The optimal radius jump between $\sigma=2000$ and $3000$m occurs when average user distance exceeds flight altitude. At this threshold, the UAV prioritizes flying closer to users rather than maintaining altitude to maximize throughput.}

\subsection{Impact of height and distance on spectral efficiency gain}
\begin{figure}
    \centering
    \includegraphics[width=0.9\linewidth]{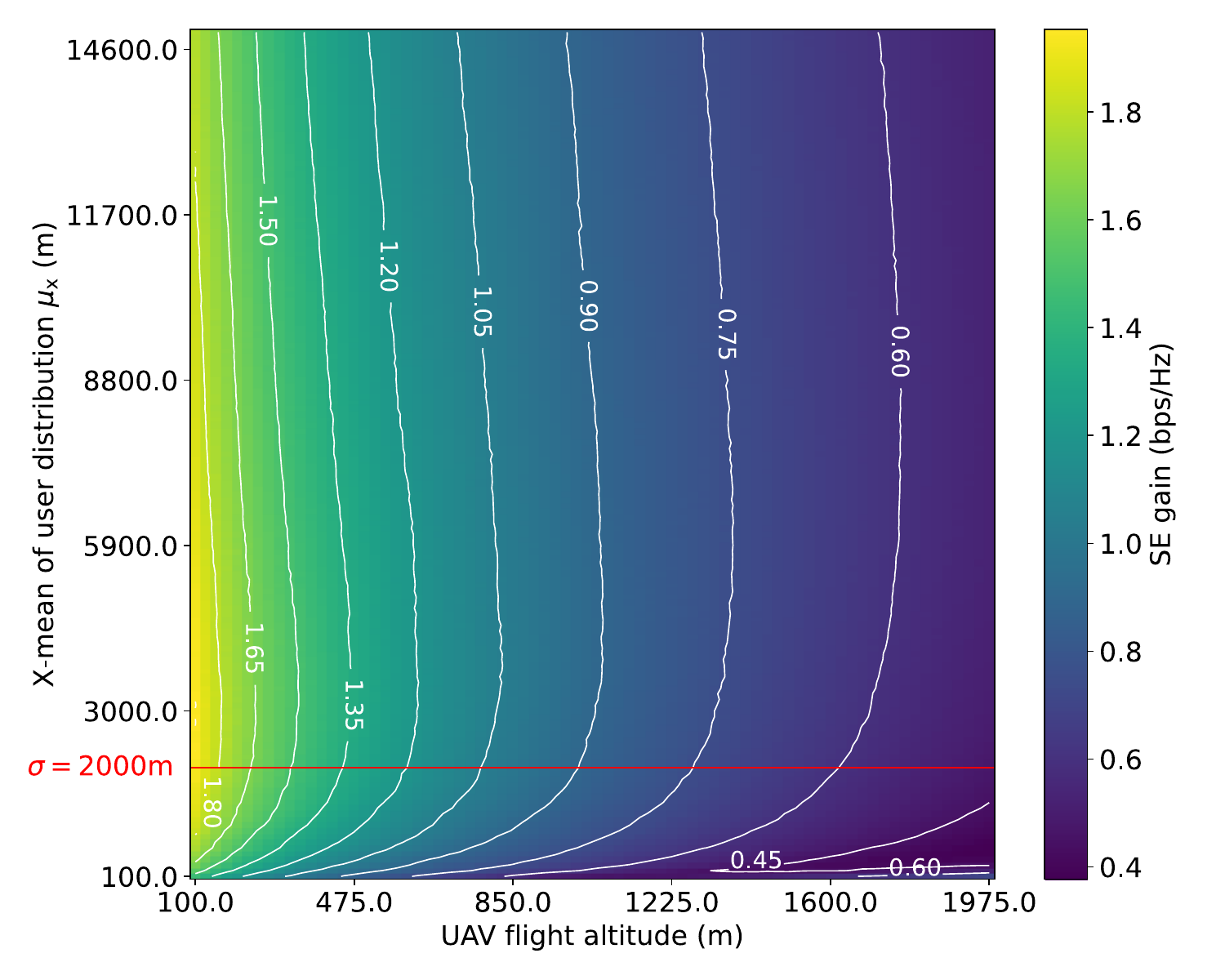}
    \caption{SE gain of a system with optimized parameters over a system with static parameters relative to the height of the UAV and the distance of the users from the BS. Flight altitude is highly correlated with SE gain because variance is increasingly impactful at lower altitudes. User distance is less so because both optimized and static models are able to account for it with timesharing.}
    \label{fig:heatmap}
\end{figure}
\topic{Lastly, we examine the joint effect of UAV height and user distance on SE gain.} We define SE gain as the SE difference between the SE achieved by our solution and the static baseline, a metric we use to measure when our optimization is most useful. The heatmap in \figref{fig:heatmap} measures SE gain relative to the flight altitude and the mean of the user distribution, with standard deviation fixed at $\distXDev=\distYDev=$ 2000m. The results show that the SE gain is highly correlated with flight altitude, but largely independent of the distribution's mean. This independence occurs because both models account for user distance similarly via timeshare biasing. In contrast, altitude has a large impact on SE gain. At high altitudes, the user distribution appears as a point source, making the effects of variance and hotspots negligible. At low altitudes, our optimization provides significant gains by adapting flight parameters to mitigate variance.

\section{Conclusions} \label{sec:conclusions}
\topic{Fixed-wing UAVs can provide quick and effective solution for disrupted connectivity over a coverage hole during disasters and emergencies}. In this paper, we proposed an achievable spectral efficiency optimization for a fixed-wing UAV relay on a circular trajectory that connects ground users to a distant BS. We showed that a fixed-wing UAV relay with circular trajectory can provide an achievable SE close to that of an unattainable upper-bound. Moreover, optimizing timeshare, trajectory, and user scheduling introduces a substantial achievable SE improvement compared to a heuristic model. Furthermore, the optimization is able to outperform the heuristic model as standard deviation increases, and does especially well at low altitudes. Lastly, the UAV transmission power is positively correlated with the optimal flight radius in the cases with greater standard deviation of users. Therefore, higher UAV transmission power for distant users is more beneficial for larger coverage holes with more user deviation. Future work will focus on extending this model to handle more advanced trajectories, and consider the benefits and challenges of using multiple UAVs in the relaying system.

\bibliographystyle{IEEEtran} 
\bibliography{./references.bib}

\end{document}